\documentclass{article}

\usepackage{arxiv}

\usepackage[utf8]{inputenc} % allow utf-8 input
\usepackage[T1]{fontenc}    % use 8-bit T1 fonts
\usepackage{hyperref}       % hyperlinks
\usepackage{url}            % simple URL typesetting
\usepackage{booktabs}       % professional-quality tables
\usepackage{amsfonts}       % blackboard math symbols
\usepackage{nicefrac}       % compact symbols for 1/2, etc.
\usepackage{microtype}      % microtypography
\usepackage{lipsum}
\usepackage{graphicx}
\usepackage{subfigure}
\title{Processing topical queries on images of historical newspaper pages}

\author{
  Jos\'e E. B. Maia and Gild\'acio J. de A. S\'a \\
  Universidade Estadual do Cear\'a – UECE \\ 60714-903 - Fortaleza - Cear\'a - Brasil \\
  \texttt{jose.maia@uece.br, gildacio.sa@gmail.com} \\
}

\begin{document}
\maketitle

\begin{abstract}
Historical newspapers are a source of research for the human and social sciences. However, these image collections are difficult to read by machine due to the low quality of the print, the lack of standardization of the pages in addition to the low quality photograph of some files. This paper presents the processing model of a topic navigation system in historical newspaper page images. The general procedure consists of four modules which are: segmentation of text sub-images and text extraction, preprocessing and representation, induced topic extraction and representation, and document viewing and retrieval interface. The algorithmic and technological approaches of each module are described and the initial test results about a collection covering a range of 28 years are presented.
\end{abstract}

% keywords can be removed

\section{Introduction}
Newspaper texts have the primacy of being contemporary to the facts, being rich in micro-details and circumstantial details to the fact reported. However, due to the urgency of its publication, the critical and contextualized analysis of events only occurs a posteriori, being the object of social study. Thus, historical newspapers preserve a rich moment of that society that must be preserved and rescued for analysis. That is why there are large collections of historical newspapers in the form of page images around the world which are of interest to anthropologists, sociologists and historians in general \cite{tumbe2019corpus,brasil2020historia,allen2010historians}. However in this way of storing they are costly and tedious to consult.

Transforming a collection of historic newspapers for digital access and consultation is challenging in many ways. The original pages on paper have been worn out due to handling and in general are dry and brittle so that the photograph of the pages is the first step most used because it is the least invasive. Due to factors such as lack of standardization, text misalignment, page wear and poor quality of photographs in some cases, the OCR (Optical Character Recognition) \cite{martinek2019training} process often returns incomplete words, noises of various types and disconnected passages.

As for the natural language processing tasks, it presents other challenges. Note that a regional newspaper collection is a closed collection and as such may have a limited vocabulary resulting in searches using a broad universal vocabulary that may not retrieve existing relevant documents.

Because the collection of texts usually covers several decades of years, the different contexts in which the same subject appears over time creates a situation of multiples contexts (multimodality of contexts) in the statistical representation of a topic. Unsupervised topical modeling such as LDA (Latent Dirichlet Allocation) \cite{blei2003latent} is useful in exploratory search however we need a search directed by the user's interest represented by a query. The researcher has a specific query that may not be frequent in the collection but whose occurrence is important and will not appear as a topic in LDA topical modeling. These observations led us to develop a technique that we call induced topics in which the user provides one or more seed terms that should guide the characterization of the topic of interest. The induced topic procedure is performed as a post-processing stage of the LDA and returns a vector of words representing the topic signature which is used to retrieve the fragments of text of interest.
In continuity, Section 2 describes the general procedure and algorithms, Section 3 is about experiments and results and the conclusion comes in Section 4.

\section{Methods}
\begin{figure}
    \centering
    \includegraphics[scale=0.7]{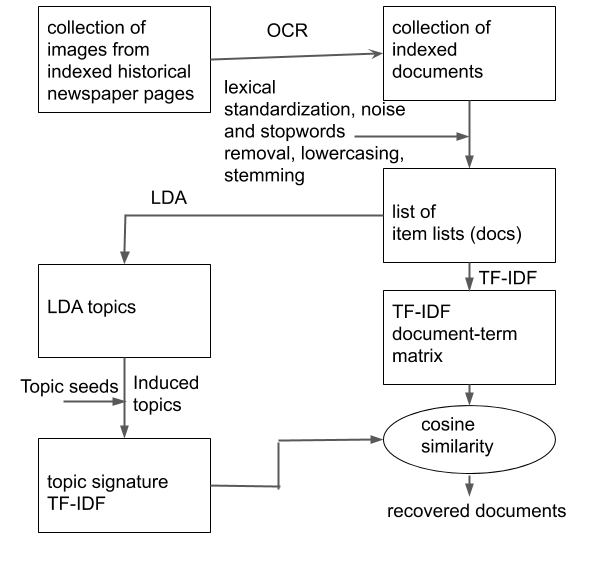}
    \caption{General procedure used in the processing for browsing topics in historical newspapers.}
    \label{fig:general-proc}
\end{figure}

\noindent \textbf{Corpus construction}: Figure \ref{fig:general-proc} shows the functional blocks and information flows of the general procedure. It is assumed that a collection of images from indexed historical newspaper pages is available from which a collection of documents indexed via OCR \cite{martinek2019training} is obtained. This is a complex step involving image segmentation, border extraction, extraction of text sub-images with different types and sizes of characters and finally the optical character recognition. The effectiveness of this phase can be as low as 60\% due to the low overall quality of the original papers from which the page images were obtained. However, the performance of the entire process ahead depends on this step. This step was supported by the commercial software AbbyyFine Reader CE \textsuperscript{\copyright} \cite{shapenko2018abbyy}.

The output of OCR faces a preprocessing step that includes lexical standardization, noises and stopwords removal, and stemming \cite{manning2008introduction}. Lexical standardization was necessary because a newspaper's database spans decades of years during which the lexicon undergoes significant changes. The output is each text document represented as a list of lexical items with the entire collection forming a list of item lists. From this list, the TF-IDF (term frequency - inverse document frequency) \cite{manning2008introduction} representation in the form of a term-document matrix is obtained for the entire collection (right branch in the Figure \ref{fig:general-proc}).

\noindent \textbf{Induced topics}: The approach here is algorithmic. The reader interested in a principled approach to labeling topics should consult \cite{arora2012learning,mimno2014low}. Initially, the number of topics of interest $N$ and one or more seed words for each topic are assumed to be available. The definition of the number of topics of interest and the seeds comes from knowledge of the context or domain of application. The search for a single induced topic is an instance.

First, the LDA algorithm \cite{blei2003latent} is applied to the corpus with the number of topics $M = n \times N$ as the input parameter. This fragmentation of LDA topics aims to model the statistical multimodality of context in the corpus. The return of the LDA is the set of words that make up each topic with their relative weights in the topic. Note that the same word can be on different topics but usually with different weights. It is possible that this unsupervised procedure returns an outline in which one or more seed words do not have a relevant weight in the LDA topics. In this case, the process is repeated with an increasing number of topics, increasing $n$, until a desired configuration is obtained.

The Induced Topic introduced here functions as a query expansion (Query Expansion) \cite{manning2008introduction} in which each topic is represented by a signature with $K$ words not repeated in other topics. To obtain them, we used the following greedy search algorithm applied to the results of the LDA: for each seed word, it searches on which topics the word is the most important. These LDA topics are labeled with that word and the process continues until you label all topics.
Then, each set of LDA topics labeled by the same seed forms an Induced Topic. Then the next most important word in each topic is taken. If there is no repetition between topics, each one is added to your group's signature. Each word assigned to a topic is subtracted from the other topics where it appears with lower weights to ensure non-repetition. If the same word has the highest weight in two or more groups, it is allocated to the group where the weight is greatest. The process is repeated until each topic is represented by $K$ words, where $K$ is a predefined value. In this work, each topic was represented by $K = 10$ words. Correction in this general process is included to handle exceptions.

\noindent \textbf{Information retrieval}: It begins by representing the signature of the topical query in TF-IDF in the space of terms in which the corpus was represented. The cosine similarity measure \cite{manning2008introduction} given by $cos(\mathbf{x},\mathbf{y}) = \frac{\mathbf{x}.\mathbf{y}}{||\mathbf{x}||.||\mathbf{y}||}$ is then used to rank documents by importance. In this equation, $\mathbf{x}$ and $\mathbf{y}$ represent the TF-IDF vectors for the topical query and the document respectively, and $\mathbf{x}.\mathbf{y}$ represents the dot product. An additional feature is that a minimum number of terms in each document can be set to filter documents of interest by size. This characteristic proved to be useful for a fragmented corpus obtained by OCR on the poor quality originals of historical newspapers.

%\begin{equation}
%   cos(\mathbf{x},\mathbf{y}) = %\frac{\mathbf{x}.\mathbf{y}}{||\mathbf{x}||.||\mathbf{y}||}
%\end{equation}.

\section{Data, Experiments and Results}

\subsection{Data set}
The corpus used for this work, in Portuguese language, was entirely built by the authors. It represents the photographic (digital) rescue of 36,617 images of pages from the \textbf{O NORDESTE} newspaper, published by the Catholic Archdiocese of Fortaleza-CE-Brazil, during the period from 1922 to 1964. We use quotes for Portuguese words.

The low quality of the paper originals, in addition to the typology worn by time, made it difficult to capture the words. To quantify these notions, on a typical page the text subimage taken at random has 532 words, OCR was able to rescue 318 words, representing almost 60\%. This percentage improved to 78\% in the last editions (1960 and 1964) and reduced to less than 50\% in the editions of the 1920s, due to the quality of the paper and the wear of the letters.

This, however, still does not mean that 60\% of the words are useful, since the lexical corrections and standardization, inverted accents, line breaks and others must still be applied. As the spelling of the time was very different, then the adjustment of the corpus was something important to be done.

For the topic `eleição' 20,684 texts were labeled and for the topic `educação' 67,282 texts were obtained, out of a total of 6167052 texts in the corpus, which represents 0.34\% and 0.11\%, respectively. This imposes a heavily unbalanced text categorization task. These numbers refer to all types of text that have been targeted. They include short or long texts, loose sentences, small advertisements or fragments of text.

\subsection{Topical queries}
\noindent \textbf{Topical query for `eleição'}: The first topical query used the seed word `eleição'. For this query, the topic signature for $K = 10$ resulted in \textit{S\textsubscript{el} = \{`eleição',`partido',`seção',`mesa',`pleito',`voto',`presidente',`chapa',`titulo',`urna'\}} and the cosine similarity threshold was experimentally adjusted to return a small number of results. Since these are highly unbalanced one-class classification experiments, only the $precision$ metric was calculated here. The other two metrics commonly used to evaluate information retrieval algorithms, Recall and F1-measure, are not useful to calculate in this context. Both are evidently very low.

Table \ref{tab:conf_mats} shows the confusion matrix obtained in this test for $cos (x, y) \geq 0.82$. The precision performance index resulted in $precision = 88/102 = 0.8627$ or 86.27 \%. On the other hand, when considering only the first 20 texts retrieved in order of relevance (top 20) as positive (total of the first column of the confusion matrix) in this test. The ground truth in this table was carried out a posteriori by reading the recovered texts. It can be seen that most of the recovered texts actually deal with the consulted topic. The top-20 precision for the `eleição' query, however, was $precision = 15/20 = 0.75$ or 75 \%.

\noindent \textbf{Topical query for `educação'}: The topic signature for $K = 10$ resulted in \textit{S\textsubscript{ed} = \{`educação',`faculdade', `instituto',`nacional',`universidade',`curso',`direito',`brasil',`trabalho',`grupo'\}}. The confusion matrix obtained in this test for $cos (x, y) \geq 0.88$ is shown in Table \ref{tab:conf_mats}. The precision performance index resulted in $precision = 222/235 = 0.9464$ or 94.64 \%. 

The first 20 texts retrieved as positive (total of the first column of the confusion matrix) were ordered decreasing in relevance. The ground truth was carried out a posteriori by reading the recovered texts. It was seen that, also in this test, most of the recovered texts actually deal with the topic consulted. The top-20 precision for the `educação' query, however, was $precision = 16/20 = 0.80$ or 80 \%.

\begin{table}[ht]
    \centering
    \begin{tabular}{c|c|c|c||c|c|c}
        \hline
         & \multicolumn{6}{c}{predicted} \\
         \hline
        & \multicolumn{2}{c}{`eleição'} & &\multicolumn{2}{c}{`educação'} & \\
         \hline
         ground truth & P & N & total & P & N & total\\
         \hline
        P' & 88 & 20,596 & 20,684 & 222 & 67,060& 67,282\\
         \hline
        N' & 14 & 6,146,354 & 6,146,368 & 13 & 6,099,979& 6,099,992\\
         \hline
        total & 102 & 6,166,950 & 6,167,052 & 235 & 6,166,817 & 6,167,052\\
         \hline
    \end{tabular}
    \vspace{3mm}
    \caption{Confusion matrices for topical queries `eleição' and `educação'.}
    \label{tab:conf_mats}
\end{table}

\section{Conclusion}
The concepts and methods adopted in the design of a System of Processing and Navigation by Topics in Images of Pages of Historical Newspapers were described and the results of a proof of concept evaluation were presented and analyzed. In addition to the contribution to the systemic project, this work proposed and preliminarily evaluated a semi-supervised approach to the problem of the generation and organization of subjects by topic.

Specifically, starting from one or more seed words per topic, the algorithm extends the topic coverage by processing the LDA output to build the topic signature which will be used as a topical query.

A third contribution of this work was to build a new data set of images of pages of a historical newspaper through the photographic (digital) rescue of 36,617 page images of the \textbf{O NORDESTE} newspaper, published by the Catholic Archdiocese of Fortaleza-CE-Brazil in the period from 1922 to 1964. The proof-of-concept evaluation produced encouraging results.

\bibliographystyle{unsrt}  
\bibliography{arxivHNbib}  %%% Remove comment to use the external .bib file (using bibtex).

\begin{thebibliography}{1}

\bibitem{tumbe2019corpus}
Chinmay Tumbe.
\newblock Corpus linguistics, newspaper archives and historical research
  methods.
\newblock {\em Journal of Management History}, 2019.

\bibitem{brasil2020historia}
Eric Brasil and Leonardo~Fernandes Nascimento.
\newblock Digital history: reflections from the brazilian digital hemerotheque
  and the use of caqdas in the re-elaboration of historical research, in
  portuguese.
\newblock {\em Revista Estudos Hist{\'o}ricos}, 33(69):196--219, 2020.

\bibitem{allen2010historians}
Robert~B Allen and Robert Sieczkiewicz.
\newblock How historians use historical newspapers.
\newblock {\em Proceedings of the American Society for Information Science and
  Technology}, 47(1):1--4, 2010.

\bibitem{martinek2019training}
Ji{\v{r}}{\'\i} Mart{\'\i}nek, Ladislav Lenc, and Pavel Kr{\'a}l.
\newblock Training strategies for ocr systems for historical documents.
\newblock In {\em IFIP International Conference on Artificial Intelligence
  Applications and Innovations}, pages 362--373. Springer, 2019.

\bibitem{blei2003latent}
David~M Blei, Andrew~Y Ng, and Michael~I Jordan.
\newblock Latent dirichlet allocation.
\newblock {\em Journal of machine Learning research}, 3(Jan):993--1022, 2003.

\bibitem{shapenko2018abbyy}
Andrey Shapenko, Vladimir Korovkin, and Benoit Leleux.
\newblock Abbyy: the digitization of language and text.
\newblock {\em Emerald Emerging Markets Case Studies}, 2018.

\bibitem{manning2008introduction}
Christopher~D Manning, Prabhakar Raghavan, and Hinrich Sch{\"u}tze.
\newblock {\em Introduction to information retrieval}.
\newblock Cambridge university press, 2008.

\bibitem{arora2012learning}
Sanjeev Arora, Rong Ge, and Ankur Moitra.
\newblock Learning topic models--going beyond svd.
\newblock In {\em 2012 IEEE 53rd Annual Symposium on Foundations of Computer
  Science}, pages 1--10. IEEE, 2012.

\bibitem{mimno2014low}
David Mimno and Moontae Lee.
\newblock Low-dimensional embeddings for interpretable anchor-based topic
  inference.
\newblock In {\em Proceedings of the 2014 Conference on Empirical Methods in
  Natural Language Processing (EMNLP)}, pages 1319--1328, 2014.

\end{thebibliography}

%\begin{thebibliography}{1}

%\bibitem{kour2014real}
%George Kour and Raid Saabne.
%\newblock Real-time segmentation of on-line handwritten arabic script.
%\newblock In {\em Frontiers in Handwriting Recognition (ICFHR), 2014 14th
%  International Conference on}, pages 417--422. IEEE, 2014.

%\bibitem{kour2014fast}
%George Kour and Raid Saabne.
%\newblock Fast classification of handwritten on-line arabic characters.
%\newblock In {\em Soft Computing and Pattern Recognition (SoCPaR), 2014 6th
%  International Conference of}, pages 312--318. IEEE, 2014.

%\bibitem{hadash2018estimate}
%Guy Hadash, Einat Kermany, Boaz Carmeli, Ofer Lavi, George Kour, and Alon
%  Jacovi.
%\newblock Estimate and replace: A novel approach to integrating deep neural
%  networks with existing applications.
%\newblock {\em arXiv preprint arXiv:1804.09028}, 2018.

%\end{thebibliography}

\end{document}